\documentclass[12pt]{article}

\usepackage{times}
\usepackage[dvips]{graphicx}

\begin{document}
\begin{titlepage}

\hfill{LIGO-970174-R}

\begin{center}

\vfill
{\Large\bf Optical vernier technique for in-situ measurement 
of the length of long Fabry-Perot cavities}
\footnote{submitted to Measurement Science and Technology}

\vspace{1cm}
{M. Rakhmanov, M. Evans and H. Yamamoto}

\vspace{1 cm}
{\it LIGO Project \\
California Institute of Technology \\
Pasadena, CA 91125} 

\end{center}

\vfill
\begin{abstract}
We propose a method for in-situ measurement of the length of kilometer size
Fabry-Perot cavities in laser gravitational wave detectors. The method is
based on the vernier, which occurs naturally when the laser incident on the
cavity has a sideband. By changing the length of the cavity over several
wavelengths we obtain a set of carrier resonances alternating with sideband
resonances. From the measurement of the separation between the carrier and a
sideband resonance we determine the length of the cavity. We apply the
technique to the measurement of the length of a Fabry-Perot cavity in the
Caltech 40m Interferometer and discuss the accuracy of the technique.
\end{abstract}

\vfill
\end{titlepage}

\section{Introduction}

Very long Fabry-Perot cavities serve as measuring devices for interferometric
gravitational wave detectors, which are currently under construction
\cite{Bradaschia:1990,Abramovici:1992,Tsubono:1995}. Among them is the Laser
Interferometer Gravitational wave Observatory (LIGO) which will have 4 km
long cavities \cite{Abramovici:1992}. The cavity length, defined as the
coating-to-coating distance between its mirrors, is an important parameter
for these gravitational wave detectors. It determines the detector
sensitivity and its overall performance. Therefore, the length must be known
with high accuracy, especially if more than one wavelength of laser is
required to resonate in the cavity. Since the length of LIGO Fabry-Perot
cavities can change by 0.4 mm due to ambient seismic motion of the ground we
do not need to measure the length with accuracy better than a millimeter.

Measurement of distances of order a few kilometers with millimeter accuracy
requires special techniques, such as GPS or optical interferometry.
Application of the GPS technique would be difficult because the mirrors of
the gravitational wave detectors are inside a vacuum envelope and the GPS
receivers cannot be placed very close to the reflective surfaces of the
mirrors. On the other hand optical interferometry provides both convenient
and precise way to measure distances \cite{Hariharan:1992}. Common
techniques are the method of fractional fringe, the method of synthetic
wavelength and the method frequency scanning \cite{Zhu:1991}. Variations of
these techniques for different applications are discussed in the SPIE
Publication \cite{Bosch:1995}. Although these techniques may provide high
precision length measurements (100 $\mu$m and better) they are not well
suited for Fabry-Perot cavities of the gravitational wave detectors. All
these techniques require installation of additional optics, modification of
the detector configuration and alignment.

In this paper we propose a technique for in-situ measurement of the cavity
length which requires no special equipment or modification to the
interferometer. The technique is based on the ability of the Fabry-Perot
cavity to resolve close spectral lines. The only requirement is that there
be at least two close wavelengths in the laser incident on the Fabry-Perot
cavity. This requirement will be easily satisfied by all gravitational wave
detectors, which are currently under construction, because optical sidebands
are an essential part of their signal extraction schemes.

For single wavelength input laser a Fabry-Perot cavity produces an array of
resonances along its optical axis. The resonances are  equally spaced and
separated by the half-wavelength of the laser. By moving one of the mirrors
over several wavelengths, and thus changing the cavity length, we can
observe these resonances. Two slightly different wavelengths give rise to
two sets of resonances with slightly different spacings, thereby forming a
vernier scale along the optical axis.

Mechanical verniers have been extensively used in various precision
measurement devices, such as calipers and micrometers. The idea of a vernier
is that an enhanced precision is obtained if two slightly different length
scales are used simultaneously \cite{Kent:1950}, \cite{Moffitt:1975}. The
technique we propose here is an extension of the vernier idea to the length
scales set by the laser wavelengths.

Our method is similar to the method developed by Vaziri and Chen
\cite{Vaziri:1997} for application to multimode optical fibres. They obtain
the intermodal beat length of the two-mode optical fibres by measuring a
separation between the resonances corresponding to these modes. We developed
our method independently of them for application to the very long
Fabry-Perot cavities in gravitational wave detectors. Although different in
motivation and underlying physics our method resembles theirs, because of the
common vernier idea.

\section{Theory of vernier method}

A mechanical vernier is a combination of two length scales which usually
differ by $10\%$. The optical vernier, described in this paper, is made out
of two laser wavelengths which differ by roughly one part in $10^8$. To use
the laser wavelengths in exactly the same way the mechanical verniers are
used would be impossible. Instead we relate the optical vernier with the beat
length, as we describe below.

Let the primary length scale be $a$ and a secondary length scale be $a'$.
Assume that $a'>a$ and consider two overlapping rulers made out of these
length scales, which start at the same point. Let $z$ be a coordinate along
the rulers with origin at the starting point. The coordinates for the two
sets of marks are
\begin{eqnarray}
   z  & = & N  a , \label{zdef} \\
   z' & = & N' a', \label{zpdef}
\end{eqnarray}
where $N$ and $N'$ are integers. Each mark on the secondary rule is shifted
with respect to the corresponding mark on the primary ruler. The shift
accumulates as we move along the $z$-axis. At some locations along $z$-axis
the shift becomes so large that the mark on the secondary ruler passes the
nearest mark on the primary ruler. The first passage occurs at $z=b$, where
$b$ is the beat length, defined according to the equation
\begin{equation}\label{b}
   \frac{1}{b} = \frac{1}{a} - \frac{1}{a'}.
\end{equation}
Other passages occur at multiples of the beat length:
\begin{equation}\label{mb}
   y = m b ,
\end{equation}
where $m$ is integer. Thus the number of beats within a given length, $z'$,
is equal to the integer part of the fraction $z'/b$. The beat number, $m$,
is related to the order numbers of the two nearest marks on the different
rulers:
\begin{equation}\label{m}
   m = N - N'.
\end{equation}
Periodicity of the beats manifests itself in the similarity relation 
\begin{equation}\label{ident}
    \frac{z' - z}{a} = \frac{z' - y}{b}.
\end{equation}
Derivation of this identity is given in the Appendix. 

Let us define the shift of the mark at $z'$ on the secondary ruler with
respect to the nearest mark at $z$ on the primary ruler as a fraction
\begin{equation}
    \mu = \frac{z' - z}{a}.
\end{equation}
Then the similarity relation, equation~(\ref{ident}), allows us to express 
the length of the secondary ruler in terms of the beat length:
\begin{eqnarray}
    z' & = & y + \mu b \label{keyEq1} \\
       & = & (m + \mu) b \label{keyEq2}.
\end{eqnarray}
Therefore, if we know the beat number, $m$, and the shift, $\mu$, we can find 
the length of the ruler. 

\begin{figure}[ht]
\begin{center}
  \caption{An example of vernier. 
  The integers are the order numbers $N$ and $N'$. The length of the 
  secondary ruler ($z' = 5 a'$) is equal to 5.6.}
  \label{vernier}
  \includegraphics[width=12cm]{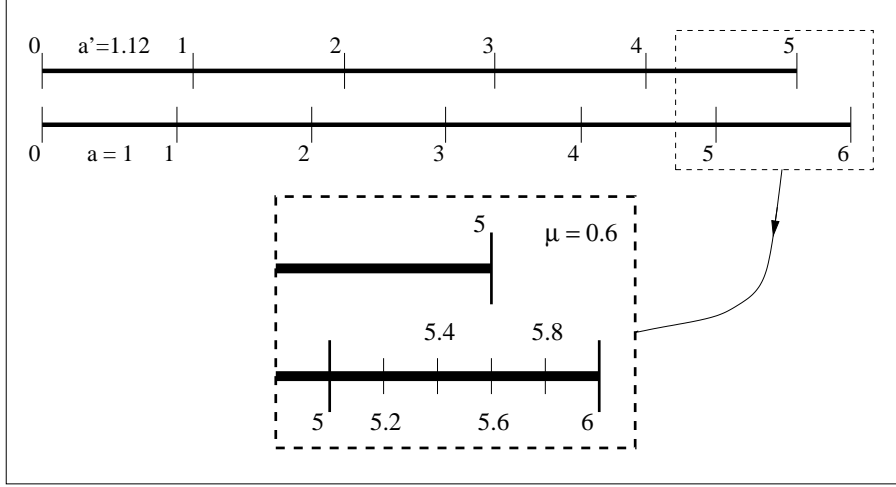}
\end{center}
\end{figure}

We illustrate the method on an example of a vernier with length scales: 
$a = 1$ and $a' = 1.12$, shown in figure~\ref{vernier}.
In this case the beat length is $9 \frac{1}{3}$.  There are no passages
within the length shown in the figure, therefore $m = 0$. From the figure
we see that the shift is equal to $0.6$. Thus we find that the length of the
secondary ruler, $z' = \mu b$, is equal to $5.6$, which is a correct result
as can be seen from the figure.

For a single wavelength laser a Fabry-Perot cavity produces an array of
resonances along its optical axis. Two slightly different wavelengths give
rise to two overlapping arrays of resonances with slightly different
spacings and form a vernier scale.

Let $z$ be a coordinate along the optical axis of the cavity. Assume that
the input mirror is placed at $z=0$ and the end mirror is at $z=L$. In the
experiment below different wavelengths are obtained by phase modulation of a
single wavelength laser. Let the frequency of the phase modulation be $f$
then the modulation wavelength is $\Lambda=c/f$. Three most prominent
components of the phase modulated laser are the carrier with wavelength
$\lambda_0$ and the first order sidebands with wavelengths $\lambda_{\pm1}$,
which are defined as
\begin{equation}
   \frac{1}{\lambda_{\pm 1}} = \frac{1}{\lambda_0} \pm \frac{1}{\Lambda}.
\end{equation}
Any two wavelengths can be used to form a vernier. For example, the primary
scale can be set by the carrier, $a=\frac{1}{2}\lambda_0$, and the secondary
scale can be set by either of the sidebands: $a'=\frac{1}{2}\lambda_{\pm 1}$.
Then the coordinates for the carrier and the sideband resonances are given
by the equations (\ref{zdef})-(\ref{zpdef}). Correspondingly, the beat
length is set by the modulation wavelength:
\begin{equation}
   b = \frac{\Lambda}{2}.
\end{equation}
Thus a vernier occurs in Fabry-Perot cavity when a multiple wavelength laser
beam is incident on it. This vernier can be used to find the cavity length. 
Similar to the length in the equation~(\ref{keyEq2}), the cavity length can 
be can be expressed in terms of the beat length:
\begin{equation}\label{L}
    L = (m + \mu) \frac{\Lambda}{2}.
\end{equation}
The beat number, $m$, can be found from the approximate length of the cavity
\begin{equation}\label{beatnum}
   m \equiv \mathrm{floor} \left( \frac{L}{\Lambda/2} \right),
\end{equation}
where ``floor'' stands for greatest integer less than. As long as the
approximate length is known with accuracy better than the beat length the
beat number is defined exactly. The shift, $\mu$, can be obtained from
observation of the carrier and sideband resonances.

\section{Measurement results and discussion}

\begin{figure}[ht]
\begin{center}
  \caption{Setup of experiment.}
  \label{setup}
  \includegraphics[width=12cm]{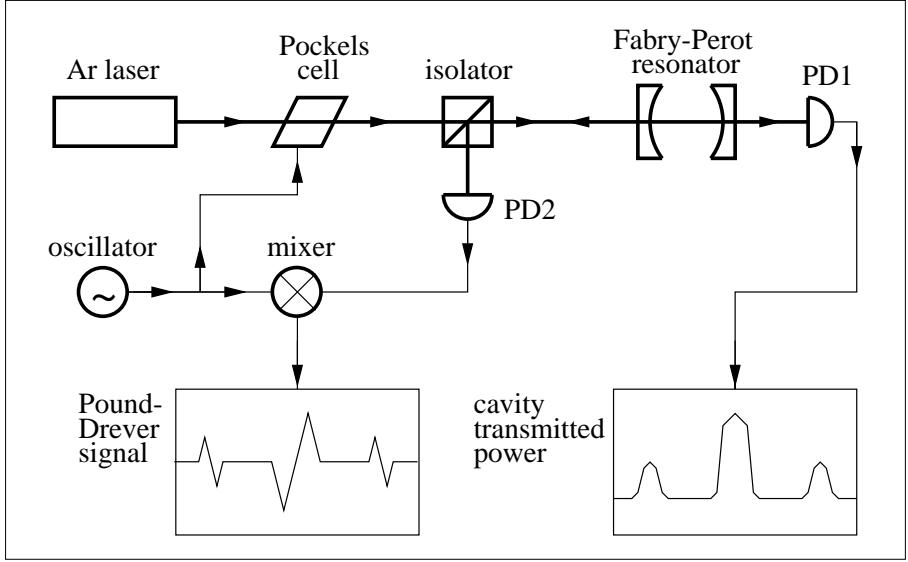}
\end{center}
\end{figure}

We apply the technique to measure the length of Fabry-Perot cavity of the 40m
prototype of LIGO interferometer at Caltech. For our measurement we use one of 
the arm cavities of the interferometer and the Pound-Drever signal extraction
scheme \cite{Drever:1983}. The setup is shown in figure~\ref{setup}.

A single wavelength ($\lambda_0=514.5$ nm) laser beam is generated by Ar 
laser. The sidebands on the laser are produced by phase modulation at the
Pockels cell, which takes its input from the RF oscillator with modulation
frequency of $32.7$ MHz. The modulation wavelength corresponding to this
frequency is $\Lambda=9.16795$ m. The resulting multi-wavelength laser beam
is incident on the Fabry-Perot cavity, whose approximate length, $L=38.5\pm
0.2$ meters, is known from previous measurements. From the approximate length
we find the beat number:
\begin{equation}
   m = 8.
\end{equation}

Both the input and the end mirror of the cavity are suspended from wires and
are free to move along the optical axis of the cavity. The signals are 
obtained from the photodiodes PD1 and PD2. The signals are: the cavity
transmitted power and the Pound-Drever signal, which is the output of the
mixer. Although either signal can be used for length measurement we choose
the Pound-Drever signal because it provides higher precision than the signal
based on the transmitted power.

\begin{figure}
\begin{center}
  \caption{Oscilloscope trace of Pound-Drever signal. The resonances 
   corresponding to the carrier and the sidebands are marked by $S_0$ 
   and $S_{\pm 1}$. Other resonances result from the higher order 
   modes due to imperfections of the laser and tilts of the mirrors.}
  \label{data}
  \includegraphics[height=8cm]{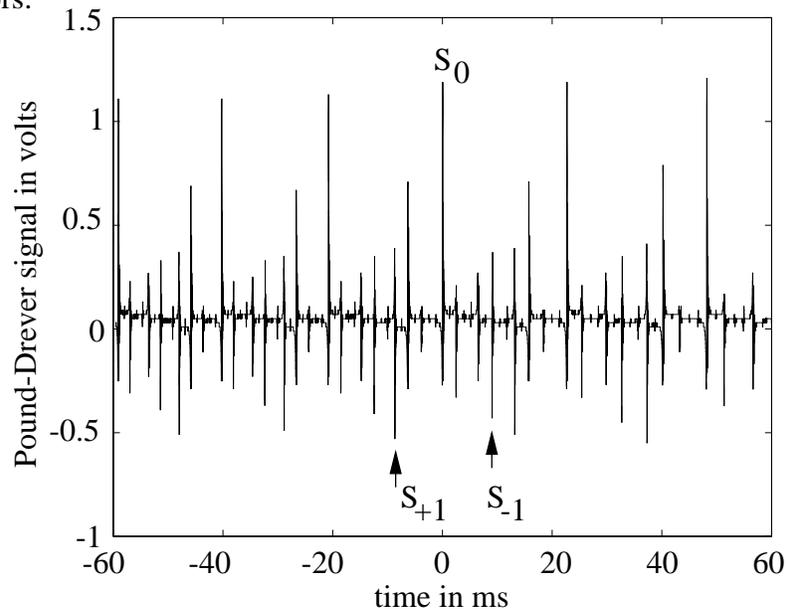}
\end{center}
\end{figure}

In the experiment the motion of the front mirror is damped by a local control
system and the end mirror is swinging freely through several wavelengths. As
the end mirror moves through the resonances sharp peaks appear in the output
signals. From the trace on the oscilloscope, figure~\ref{data}, we obtain the
times when the mirror passes through the carrier resonances, $t_0(p)$, and
the sideband  resonances, $t_{\pm 1}(p)$, where $p$ is integer from 1 to 6.
The times are found with a precision of 1 $\mu$s, set by the resolution of
the oscilloscope. 

\begin{figure}
\begin{center}
  \caption{Interpolated mirror trajectory within the first 6 carrier 
  resonances.}
  \label{traject}
  \includegraphics[height=8cm]{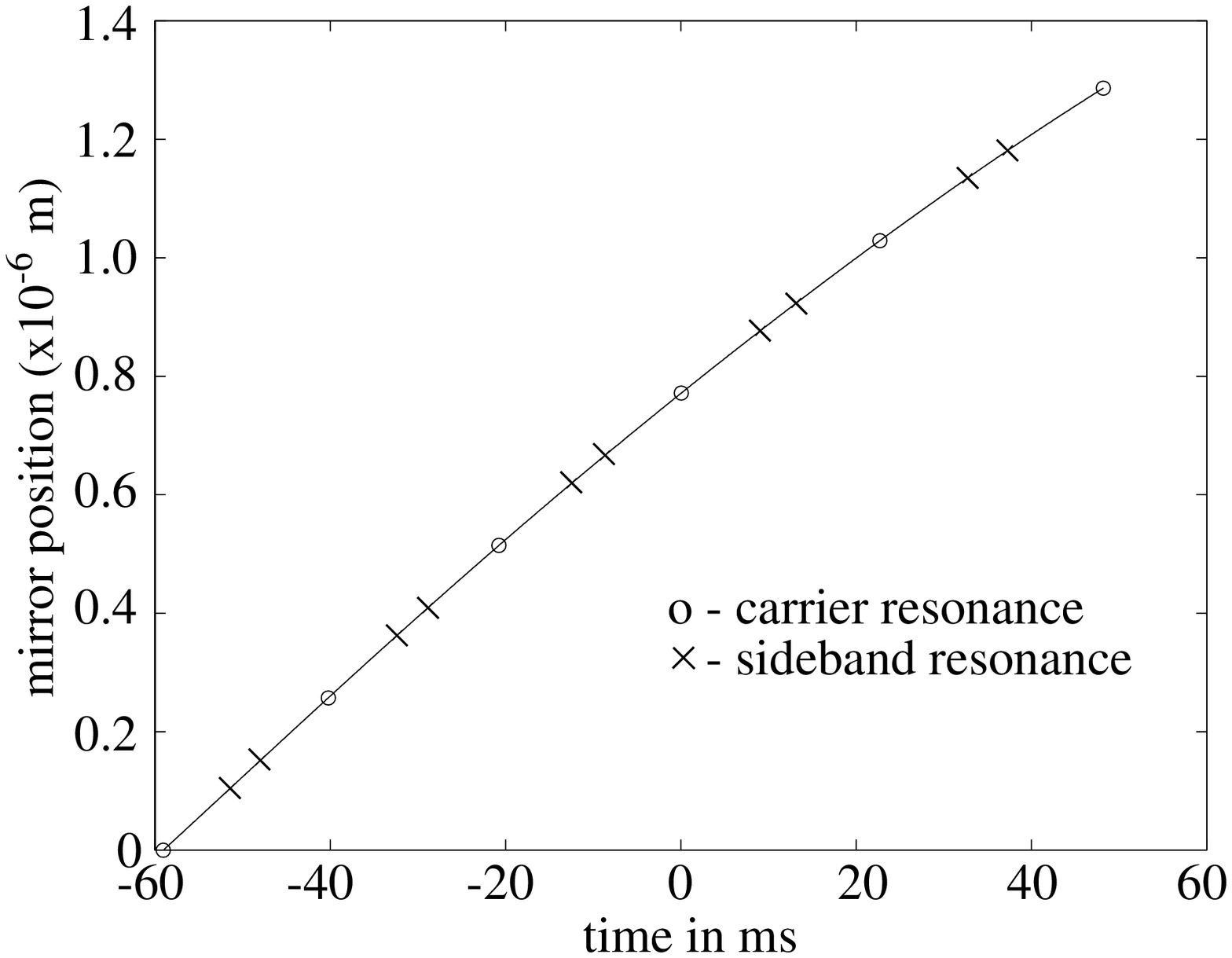}
\end{center}
\end{figure}

The carrier resonances are located at
\begin{equation}
   z_0(p) = (p - 1) \frac{\lambda_0}{2} + u,
\end{equation}
where $u$ is an unknown constant, which cancels in the calculation. The
location of the sideband resonances can be found from the times $t_{\pm1}(p)$ 
if the trajectory of the mirror is known. We find the approximate trajectory 
of the mirror by polynomial interpolation between the carrier resonances. The 
plot of the interpolated mirror trajectory is shown in figure~\ref{traject}.
Let the interpolation polynomial be $F(t)$. Using the polynomial we find the
location of the sideband resonances as follows
\begin{eqnarray}
   z_{-1}(p) & = & F(t_{-1}(p)) + u,\\
   z_{+1}(p) & = & F(t_{+1}(p)) + u.
\end{eqnarray}
Once the locations of the carrier and the sideband resonances are known we
can find the corresponding shifts as
\begin{equation}
   \mu(p) = \frac{z_{-1}(p) - z_0(p)}{\lambda_0/2}
\end{equation}
for the lower sideband, and 
\begin{equation}
   \mu(p) = \frac{z_0(p) - z_{+1}(p)}{\lambda_0/2}
\end{equation}
for the upper sideband. The results are shown in the table~\ref{results}.
\begin{table}[ht]
   \caption{The shifts obtained from the interpolated mirror trajectory.
   The first and the last fringe contains only one sideband resonance.}
   \label{results}
   \begin{center}
   \begin{tabular}{ccc}
   \hline
   \hline
   $p$ & $\mu$  & $\mu$ \\
   resonance order & ($-1$ sideband) & ($+1$ sideband) \\ 
   \hline
   1   &  0.407213  &  ...      \\
   2   &  0.409232  &  0.410154 \\
   3   &  0.408725  &  0.408647 \\
   4   &  0.408816  &  0.409038 \\
   5   &  0.409685  &  0.409093 \\
   6   &  ...       &  0.408188 \\  
   \hline
   \hline
   \end{tabular}
   \end{center}
\end{table}
The average shift and its standard deviation is  
\begin{equation}
   \mu = 0.4089 \pm 0.0008.
\end{equation}
Using the equation~(\ref{L}) we find the length of the cavity 
\begin{equation}
   L = 38 546 \pm 4 \; \mathrm{mm}.
\end{equation}

The error in the cavity length comes from the error in the beat length and
the error in the shift. In our experiment the dominant was the error in the
shift, which is mostly the error of the polynomial interpolation. The
interpolation error can be greatly reduced if the change in the cavity length
is known with high precision. This can be done, for example, by controlling
the cavity mirrors at low frequencies.

The limiting precision of the technique, $\delta L$, is determined by the
signal used to obtain the shift $\mu$. For the transmitted power the limit
comes from the finite width of resonances in the Fabry-Perot cavity. A
separation between the resonances in the transmitted power can be measured up
to a width of a resonance. Therefore,
\begin{equation}
   \delta L \sim \frac{\Lambda/2}{\mathrm{Finesse}},
\end{equation}
which is roughly 4 mm for our experiment. This precision limit does not
depend on the length of the cavity.

There is no limit due to the finite width if the resonances are observed in
the Pound-Drever signal. In this case the separation between the resonances
are found from zero-crossings or peaks in the Pound-Drever signal and the
shifts can be measured with a precision far better than the width of a
resonance. For the Pound-Drever signal the limit on the precision is given by
the uncertainty in the beat length
\begin{equation}
   \frac{\delta L}{L} \sim \frac{\delta \Lambda}{\Lambda},
\end{equation}
which is defined by stability of the oscillator. In our case 1 Hz-stability
of the oscillator sets the limit of 1 $\mu$m to the precision of the
technique.

There are two small but noteworthy systematic errors in this method: one is
due to the phase change upon reflection off the mirrors, the other is due to
the Guoy phase of the Gauss - Hermite modes of Fabry-Perot cavity
\cite{Siegman:1986}. If the phase of the reflected laser is not exactly
opposite to the phase of the incident laser at the mirror surface the
resonances in the cavity become shifted. This effect can be as large as
$\lambda/4$ per mirror and is far below the precision of the technique. The
Guoy phase also affects the location of the resonances and can be at most
$\pi/2$ for the lowest mode of the cavity. Thus the largest contribution due
to the Guoy phase is $\lambda/4$ and can also be neglected.

\section{Conclusion}

The method of optical vernier is a simple and accurate way to measure the
cavity length of the laser gravitational wave detectors in-situ. The method
requires no special equipment or modification to the detector. We tested the
method on the 40m prototype of the LIGO interferometers and attained a
precision of 4 mm. The ultimate precision of the method is defined by the
uncertainty in the beat length, and is of order a few microns. The method is
general and can be used for length measurement of any Fabry-Perot cavity,
which allows for small adjustment of its length.

\section*{Acknowledgments}

We are very grateful to G. Hu for setting up the data acquisition for the
experiment. We also thank S. Whitcomb, A. Lazzarini, J. Camp and A. Arodzero
for the discussions and the comments on the paper. This work is supported by
National Science Foundation under Cooperative Agreement PHY-9210038.

\section*{Appendix}

Derivation of the equation~(\ref{ident}) is straightforward:
\begin{eqnarray}
   z' - z & = & N'a' - N a \\
          & = & N' (a' - a) - (N - N') a \\
          & = & N' \frac{a a'}{b} - m a .
\end{eqnarray}
Dividing both sides of this equation by $a$ we obtain the similarity
relation, equation~(\ref{ident}).

\end{document}